\documentclass[showpacs,amsmath,amssymb,aps,twocolumn,prl]{revtex4-1}%,superscriptaddress
\usepackage{graphicx}
\usepackage{soul}
\usepackage[colorlinks=true,citecolor=blue,linkcolor=magenta]{hyperref}
\usepackage[usenames]{color}
\usepackage[cspex,bbgreekl]{mathbbol}
\usepackage{relsize}
\usepackage{hyperref}

\newcommand{\ket}[1]{\left| #1 \right>} % for Dirac bras
\newcommand{\horvbket}{| \raisebox{-1.5 pt}{\includegraphics[scale=0.04]{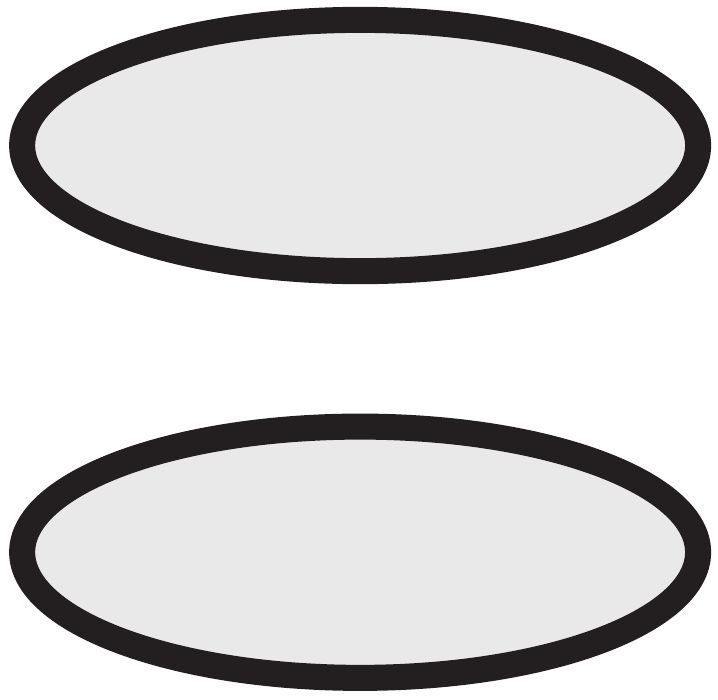}} \rangle}
\newcommand{\horsingket}{| \raisebox{-2 pt}{\includegraphics[scale=0.04]{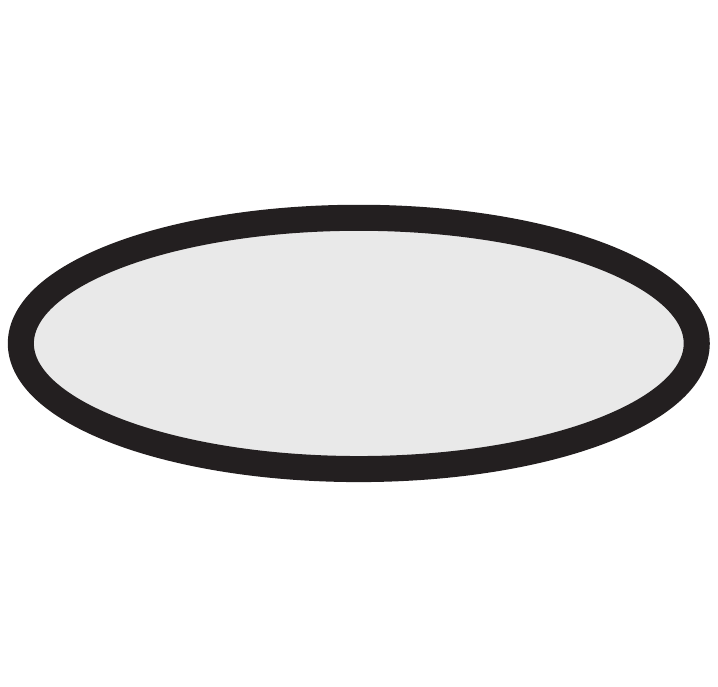}} \rangle}
\newcommand{\hortripket}{| \raisebox{-2 pt}{\includegraphics[scale=0.04]{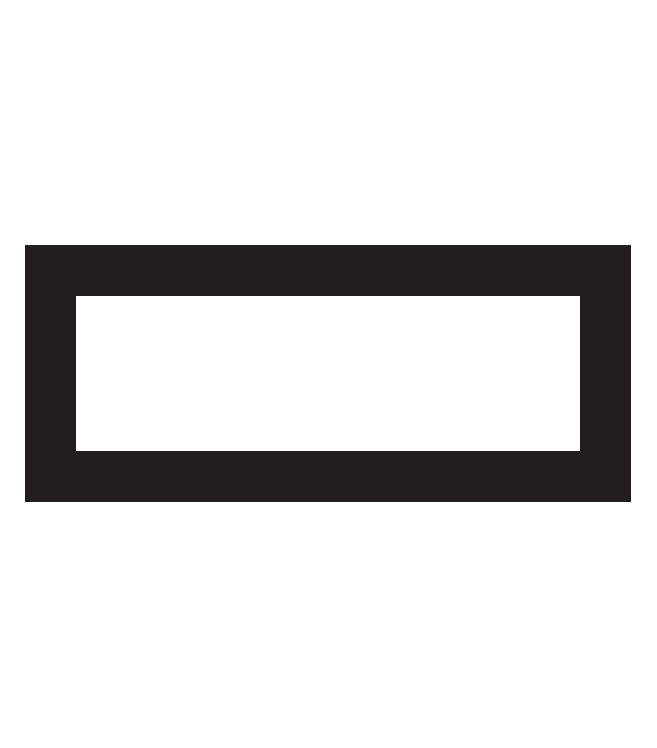}} \rangle}
\newcommand{\vertvbket}{| \raisebox{-1.5 pt}{\includegraphics[scale=0.042]{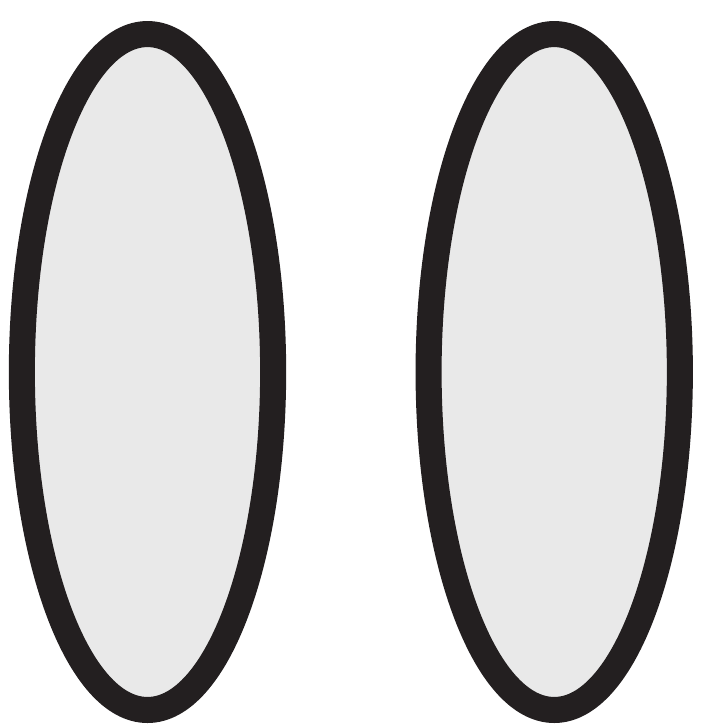}} \rangle}

\newcommand{\diagvbket}{| \raisebox{-1.5 pt}{\includegraphics[scale=0.05]{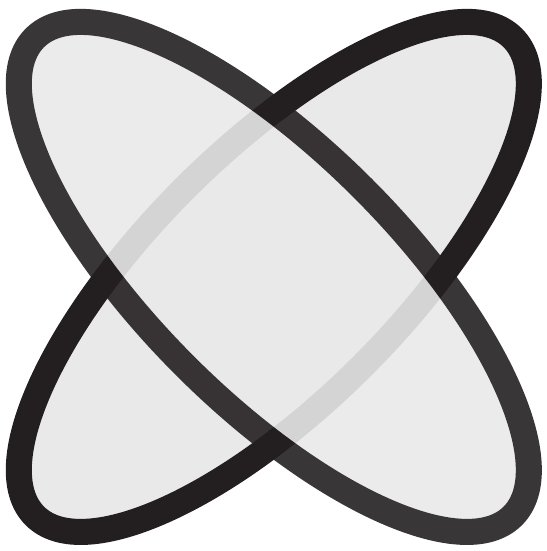}} \rangle}

\newcommand{\overlapsinglet}{\langle \raisebox{-1.5 pt}{\includegraphics[scale=0.04]{HorizontalVB.pdf}} |\raisebox{-1.5 pt}{\includegraphics[scale=0.04]{VerticalVB.pdf}} \rangle}

\graphicspath{{../figures/}}

\begin{document}

\title{Experimental realization of plaquette resonating valence bond states \\
with ultracold atoms in optical superlattices}

\author{S. Nascimb\`ene$^{1,2,3}$, Y.-A. Chen$^{1,2}$, M. Atala$^{1,2}$, M. Aidelsburger$^{1,2}$,  S. Trotzky$^{1,2}$, B. Paredes$^{4}$, and I. Bloch$^{1,2}$}

\affiliation{$^{1}$\,Fakult\"at f\"ur Physik, Ludwig-Maximilians-Universit\"at, Schellingstrasse 4, 80799 M\"unchen, Germany\\
$^{2}$\,Max-Planck-Institut f\"ur Quantenoptik, Hans-Kopfermann-Strasse 1, 85748 Garching, Germany\\
$^{3}$\,Laboratoire Kastler Brossel, CNRS, UPMC, Ecole Normale Sup\'erieure, 24 rue Lhomond, 75005 Paris, France\\
$^{4}$\,Department f\"ur Physik, Arnold Sommerfeld Center for Theoretical Physics, Ludwig-Maximilians-Universit\"at, 80333 M\"unchen, Germany}

\date{\today}

\pacs{03.75.Lm, 03.65.Xp, 75.10.Jm, 75.10.Kt}

\begin{abstract} 
The concept of valence bond resonance plays a fundamental role in the theory of the chemical bond and is believed to lie at the heart of many-body quantum physical phenomena.  Here we show direct experimental evidence of a time-resolved valence bond quantum resonance with ultracold bosonic atoms in an optical lattice. By means of a superlattice structure we create a three-dimensional array of independent four-site plaquettes, which we can fully control and manipulate in parallel. Moreover, we show how small-scale plaquette resonating valence bond (RVB) states with $s$- and $d$-wave symmetry can be created and characterized. We anticipate our findings to open the path towards the creation and analysis of many-body RVB states in ultracold atomic gases.
\end{abstract}

\maketitle
In his theory of the chemical bond, Pauling developed the concept of quantum resonance: a quantum superposition of resonant structures with different arrangements of the valence bonds \cite{pauling1931nature}. Such resonant states are essential to explain the chemical properties of certain organic molecules like benzene \cite{hueckel1931quantentheoretische}. In the context of high temperature superconductivity, Anderson extended Pauling's notion to a macroscopic level, by proposing  that electrons in Mott insulating solid state materials could form resonating valence bond (RVB) states \cite{anderson1973resonating,anderson1987resonating}. In a Mott insulating phase, electrons are localized to individual atoms or molecules, and the fluctuations in the charge (density) degree of freedom are strongly suppressed. The physics is dictated by the remaining spins, which interact via superexchange interactions. Under certain conditions, the localized spins are expected to evade local order and continue to fluctuate down to zero temperature, forming a coherent superposition of many different arrangements in which the spins are paired up into singlets or valence bonds.

Ultracold atomic gases in optical lattices \cite{jaksch2005,lewenstein2007ultracold,bloch2008many} and other quantum optical systems are promising candidates for the quantum simulation of RVB states \cite{ma2011quantum}. Their realization would allow one to gain valuable insight into the entanglement properties of these states as well as to answer fundamental questions in condensed matter physics like their stability under specific Hamiltonians such as the Hubbard model, or to test their exotic superconducting properties upon doping \cite{altman2002plaquette,trebst2006d}. In this work we create an array of small-scale versions of Pauling-like RVB states in four-site plaquettes and study their basic physical properties. Our techniques can be directly generalized to a gas of fermionic atoms, for which one expects that the adiabatic connection of such isolated plaquette RVB states could lead to the creation of a macroscopic $d$-wave superfluid state \cite{altman2002plaquette,trebst2006d,rey2009controlled}.

%
%%%%%%%%%%%%%%%%%%%%%%%%%%%%%%%%%%%%%%%%%%%%%%%%%%%%%%%%%%%%%
%   Figure Scheme
%%%%%%%%%%%%%%%%%%%%%%%%%%%%%%%%%%%%%%%%%%%%%%%%%%%%%%%%%%%%%
\begin{figure}[t!]
\includegraphics[width=\linewidth]{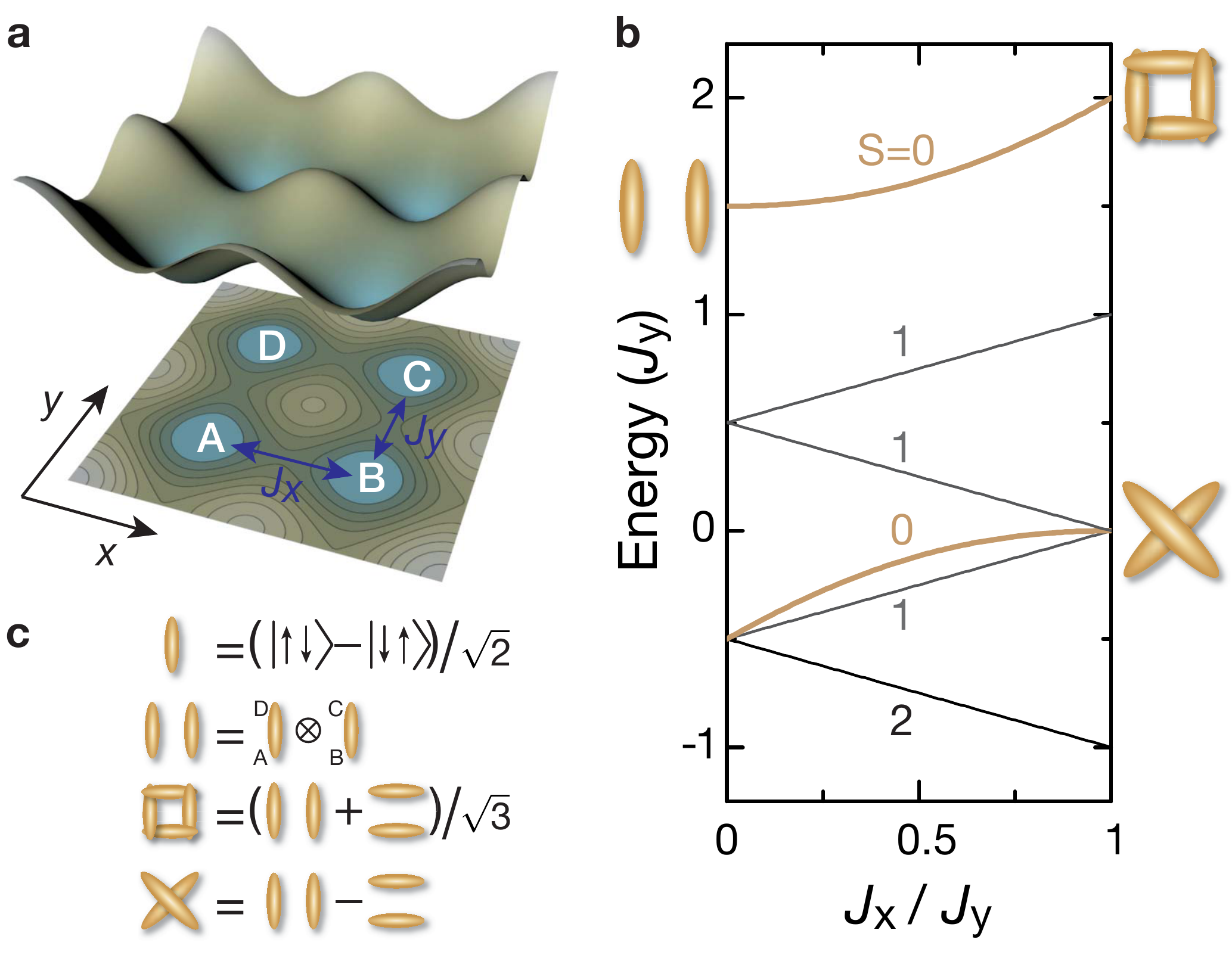}
\vspace{-0.5cm} 
\caption[]{Schematics of a single plaquette and energy levels at half filling.
(a) Scheme of the lattice potential in the $x,y$ plane, created by a pair of bichromatic optical lattices. The elementary cell is made of four wells arranged in a square configuration. (b) Energy levels of four atoms on a plaquette in a Mott insulating state at half filling, with superexchange spin couplings along $x$ $(y)$ denoted by $J_x (J_y)$. For any ratio $J_x/J_y$, the highest energy state is a total spin-$\frac{1}{2}$ singlet. In the case of $J_x/J_y=0$, it corresponds to the valence bond state $\vertvbket$, whereas for $J_x/J_y=1$ it is the $s$-wave RVB state $\left|\Phi_+\right>$. The other total singlet  for $J_x=J_y$, lower in energy, is the $d$-wave RVB state $\left|\Phi_-\right>=\diagvbket$. (c) Symbols used for a singlet bond and for the $s$-wave and $d$-wave plaquette RVB states.
\label{Fig_Scheme}} 
\end{figure}

Let us consider an ultracold gas of bosonic atoms in two internal states, loaded into a two-dimensional superlattice structure whose elementary cell is a plaquette made out of four wells arranged in a square pattern [Fig.~\ref{Fig_Scheme}(a)]. In the regime in which the tunneling amplitude between adjacent plaquettes is strongly suppressed, the system can be regarded as a collection of independent replicas of a single plaquette, the object of our study. At half filling, and when the on-site interaction $U$ dominates over the tunneling amplitude $t$ between wells in a plaquette, atoms are site localized, one per site, and the physics is governed by the remaining four effective 
$\frac{1}{2}$-spins, which interact with their next neighbors via a ferromagnetic Heisenberg interaction $J\vec{S}_i \! \cdot \! \vec{S}_j$, with $J=-4t^2/U$ \cite{duan2003controlling,garcia2003spin,kuklov2003counterflow,altman2003phase,trotzky2008time}.

To gain insight into the RVB states on a plaquette, it is convenient to write the Heisenberg interaction in terms of the swap operator $\hat{X}_{ij}=2\vec{S}_i \! \cdot \! \vec{S}_j+1/2$, a unitary operator that exchanges the states of the spins on the sites $i$ and $j$. The plaquette Hamiltonian then takes the form \cite{paredes2008minimum}:

\begin{equation}
\hat H=J_x	\hat{X}_{x}  + J_y	\hat{X}_{y},
\label{Hamiltonian}
\end{equation}

where $\hat{X}_{x(y)}$ involves exchanges of two spins along an $x$ $(y)$-bond: $\hat{X}_{x}=(\hat{X}_{AB}+\hat{X}_{CD})/2$, $\hat{X}_{y}=(\hat{X}_{AD}+\hat{X}_{BC})/2$, with $A,B,C,D$ labeling the four sites of the plaquette [Fig.~\ref{Fig_Scheme}(a)]. From now on, we consider solely the subspace of total spin zero, where all spins are part of a singlet state or valence bond. This subspace is generated by two states, which correspond to arrangements in either vertical $\vertvbket$ or horizontal bonds $\horvbket$ [Fig.~\ref{Fig_Scheme}(b)]. 
%
%%%%%%%%%%%%%%%%%%%%%%%%%%%%%%%%%%%%%%%%%%%%%%%%%
%  Fig VB oscillation
%%%%%%%%%%%%%%%%%%%%%%%%%%%%%%%%%%%%%%%%%%%%%%%%%
%
\begin{figure}[t!]
\includegraphics[width=\linewidth]{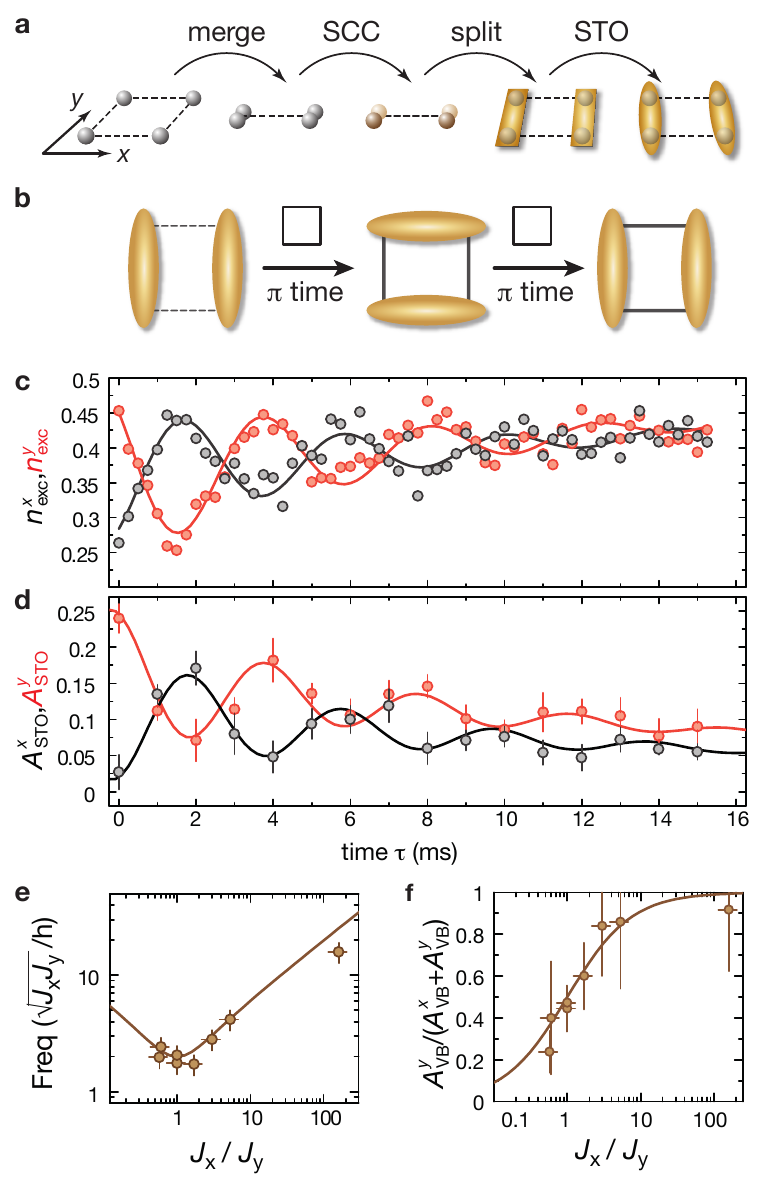}
\vspace{-0.5cm} 
\caption[bla]{Initial state preparation and valence bond oscillations. 
(a) Schematics of the preparation of an array of valence bond states $\vertvbket$ from a unit-filling Mott insulator.(b) Schematics of the valence bond oscillation: starting from $\vertvbket$, we switch on identical superexchange couplings along $x$ and $y$, leading to a coherent oscillation between $\vertvbket$ and $\horvbket$.
(c,d) Fraction of band excitations $n^{x,y}_{\mathrm{exc}}$ (c) and STO amplitude $A^{x,y}_{\mathrm{STO}}$ (d) as a function of the hold time $\tau$, with $J_x\simeq J_y=\mathrm{h}\times 120(10)$\,Hz. 
%The lattice depths were $V_{xs}=V_{ys}=12\,E_r^s$, $V_{xl}=V_{yl}=30\,E_r^l$, and $V_{z}=40\,E_r^z$.
(e) Frequency of the valence bond oscillation as a function of $J_x/J_y$.
(f) Ratio $A_\mathrm{VB}^y/(A_\mathrm{VB}^x+A_\mathrm{VB}^y)$ as a function of $J_x/J_y$, where $A_\mathrm{VB}^{x,y}$ is the initial amplitude of the valence bond oscillations as shown in (c).  The solid lines in (e) and (f) are calculated from Eq.\,(\ref{Hamiltonian}). The horizontal error bars represent the uncertainties in lattice depths and the vertical ones represent the 1$\sigma$ uncertainties of the fits to the STO traces.}
\label{Fig_VB_oscillations} 
\end{figure}

Within this subspace and for identical superexchange couplings $J_{x}=J_{y}\equiv J$, the Hamiltonian of Eq.\,(\ref{Hamiltonian}) reduces to $\hat H=-J\hat{X}_{xy}$, where $\hat{X}_{xy}=(\hat{X}_{AC}+\hat{X}_{BD})/2$ swaps two spins along a diagonal. As can directly be seen, this diagonal exchange is equivalent to a 90-degree rotation of the plaquette and converts the state $\vertvbket$ into $\horvbket$ and vice-versa, giving rise to a resonance. The eigenstates are then coherent superpositions of the form:
\[
\ket{\Phi_\pm} \propto \vertvbket \pm \horvbket.
\]
These minimum instances of RVB states exhibit no local magnetic order, and can not be distinguished from each other by measuring single-site spin observables. However they are distinct with respect to an exchange of two spins along a diagonal: the $s$-wave RVB state $\left|\Phi_+\right>$ is symmetric; the $d$-wave RVB state $\left|\Phi_-\right>$ is antisymmetric, owing to its singlet structure along the diagonals of the plaquette, $\ket{\Phi_-}=\diagvbket$. 

Our experiments began with a quasi-pure Bose-Einstein condensate of about $5\times 10^4$ $^{87}$Rb atoms in the Zeeman state $\left|F=1,m_F=-1\right>$. The atoms were loaded into a tetragonal optical lattice potential, formed by three mutually orthogonal standing waves with wavelengths $\lambda_{s}=767$\,nm (\textquotedblleft short lattices\textquotedblright) along $x$ and $y$, and $\lambda_{z}=844$\,nm along $z$. Two additional standing waves with wavelengths of $\lambda_{l}=1534\,$nm (``long lattices'') that were superimposed with the short lattices \cite{foelling2007direct} along $x$ and $y$ were then used to realize a three-dimensional periodic potential whose elementary cell is a plaquette [Fig.~\ref{Fig_Scheme}(a)]. The final lattice depths were chosen to access the Mott insulating regime with at most one atom per lattice site for our total particle number. We then employed a sequence of site merging, spin changing collision (SCC) \cite{widera2005coherent} and singlet-triplet oscillation (STO) operations \cite{paredes2008minimum,trotzky2010controlling,STO} on plaquettes [see Fig.~\ref{Fig_VB_oscillations}(a)] in order to create the initial state $\vertvbket$ out of the atomic spin states $\left|F=1,m_F=-1\right>$ and $\left|F=1,m_F=+1\right>$ \cite{SOM}.  In total we operate in parallel over about $10^3$ identical plaquettes with unit atom filling. Lattice depths of $V_{xl}=V_{yl}=35\,E_r^l$ and  $V_{z}=40\,E_r^z$ ensure negligible atom tunneling between plaquettes \cite{recoil}. 

To directly observe the valence bond resonance, the initial state $\vertvbket$ was evolved under the Hamiltonian of Eq.\,(\ref{Hamiltonian}) with identical superexchange couplings along $x$ and $y$. To this aim we ramped down the short-lattice depths in 200\,$\mu$s to $V_{xs}=V_{ys}=12\,E_r^s$, resulting in equal couplings $J_x\simeq J_y=\mathrm{h}\times 120(10)$\,Hz and a suppression of first order tunneling as $t/U\simeq1/8$. Since $\hat{X}_{xy}^2=1$, the evolved quantum state at time $\tau$ is 
\begin{equation}\label{Eq_VB}
\ket{\Psi(\tau)}=\cos\frac{\omega \tau}{2} \, \vertvbket	
-i \sin\frac{\omega \tau}{2} \, \horvbket,
\end{equation}
oscillating between the states $\vertvbket$ and $\horvbket$ with frequency $\omega=2J_{x(y)}/\hbar$. 

To characterize this state evolution, we measured the projections onto the two valence bond states: 
$\mathcal{C}_{x}=\left| \langle \Psi(\tau) \horvbket \right| ^2$, and $\mathcal{C}_{y}=\left| \langle \Psi(\tau) \vertvbket \right| ^2$, which are expected to show oscillations of amplitude $3/4$, since $\overlapsinglet=1/2$.
Within the subspace of total singlets, the observable $\mathcal{C}_x$ can be obtained either by measuring the fraction of band excitations $n_\mathrm{exc}^x=\mathcal{C}_x/2$ after merging pairs of wells along the $x$ direction, or by measuring the amplitude $A_\mathrm{STO}^x$ of STO \cite{paredes2008minimum,trotzky2010controlling,STO} induced by a magnetic-field gradient along $x$ \cite{SOM}. 
As shown in Fig.~\ref{Fig_VB_oscillations}(c) and (d), we indeed observed a coherent evolution of both $n_\mathrm{exc}^{x,y}$ and $A_\mathrm{STO}^{x,y}$. This dynamics corresponds to anti-correlated oscillations of the projections $\mathcal{C}_x$ and $\mathcal{C}_y$ that reveal the periodic swapping of the valence-bond direction.  The measured oscillation frequency $\omega/2\pi=250(10)$\,Hz is compatible with twice the value of the superexchange couplings, in agreement with Eq.\,(\ref{Eq_VB}). While the damping of the valence bond oscillation ($1/e$ decay time of 6(1)\,ms) could be attributed to inhomogeneities of the different plaquette parameters across the atomic sample, the slow overall increase of $n_\mathrm{exc}^{x}$ and $n_\mathrm{exc}^{y}$ could be caused by decoherence within a plaquette. We provide further evidence of the valence-bond dynamics governed by superexchange interactions by studying the dynamics for anisotropic couplings $J_x\neq J_y$. As shown in Fig.~\ref{Fig_VB_oscillations}(e) and (f), the measured oscillation frequencies and amplitudes as a function of $J_x/J_y$ agree well with the values predicted from the Hamiltonian dynamics of Eq.\,(\ref{Hamiltonian}).  Site-resolved population measurements were used to check that throughout the evolution the four plaquette sites remained equally populated \cite{foelling2007direct,SOM}. In the absence of residual magnetic field gradients we expect the atoms to remain in the singlet subspace $S=0$. This was checked by holding singlet atom pairs after the initial state preparation, and observing no conversion to triplet pairs.

\begin{figure}[t!]
\includegraphics[width=0.8\linewidth]{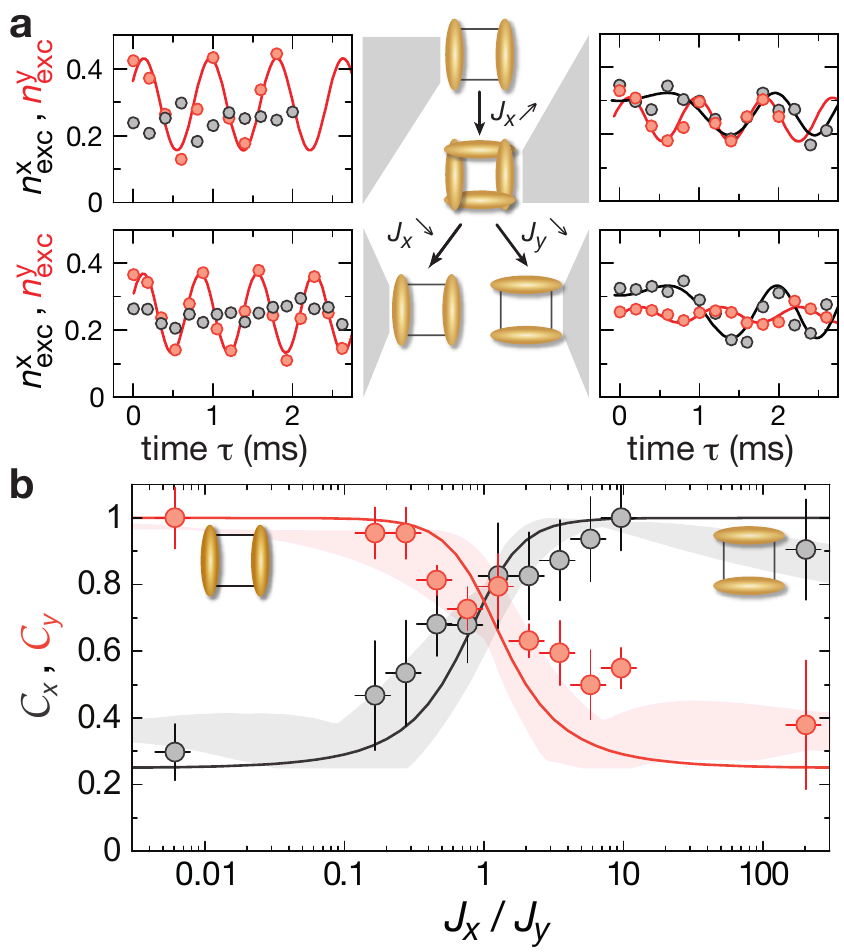}
\vspace{-0.5cm}  
\caption[]{Preparation of the $s$-wave RVB state and adiabatic valence bond swap. 
(a) Scheme of the adiabatic conversion $\vertvbket\rightarrow\ket{\Phi_+}\rightarrow\vertvbket\;\mathrm{or}\;\horvbket$, together with STO for each state. The STO period along $x$ is not constant due to an increasing magnetic gradient field during the measurements that was caused by a finite response time of the coils producing the magnetic-field gradient along $x$. The solid lines are fits of the STO taking into account the finite rise time of the magnetic field gradient for the $x$ direction.
(b) Projections $\mathcal{C}_x$, $\mathcal{C}_y$ on the valence bond states as a function of the ratio $J_x/J_y$ of superexchange couplings, measured from the STO amplitudes.   The latter were rescaled in order to give the expected value of 0.5 for the valence bond states $\horvbket$ and $\vertvbket$, using the data points at $J_x/J_y=0.006(2)$ and  $J_x/J_y=10(3)$. For the point at $J_x/J_y=200(50)$ the rate of change of the couplings was the largest and adiabaticity was not maintained. The horizontal error bars represent the uncertainties in lattice depths and the vertical ones represent the 1$\sigma$ uncertainties of the fits to the STO traces. The solid lines are calculated from the eigenstates of the Hamiltonian in Eq.\,(\ref{Hamiltonian}). The shaded lines are calculated by modeling the experimental ramps using the Schr\"odinger equation. Their widths represent the uncertainties in the lattice calibration. 
\label{Fig_Phi_plus}} 
\end{figure}

In order to create the $s$-wave RVB state $\ket{\Phi_+}$, we made use of the fact that it is adiabatically connected to the initial state $\vertvbket$ [Fig.~\ref{Fig_Scheme}(b)]. To follow this adiabatic path we started from a situation in which $V_{xs}=22 \, E_r^s$ and $V_{ys}=12 \, E_r^s$. For these parameters, $J_x/J_y$ is negligible and $\vertvbket$ is an eigenstate of the Hamiltonian in Eq.\,(\ref{Hamiltonian}). We then decreased $V_{xs}$ to 12\,$E_r^s$ within 5\,ms using an exponential ramp, converting the initial state into the $s$-wave RVB state. In order to check the adiabaticity of the lattice-depth ramps, we then increased the short lattice along $x$ ($y$) to 22\,$E_r^s$ in 5\,ms , transforming the RVB state back into a valence-bond state $\vertvbket$ (or $\horvbket$, respectively). By using STO we measured the singlet correlations along both directions $x$ and $y$ for the initial, intermediate and final states  of the ramp $\vertvbket\rightarrow\ket{\Phi_+}\rightarrow\vertvbket$ or $\horvbket$ [Fig.~\ref{Fig_Phi_plus}(a)]. As expected, for the initial state we observe oscillations close to maximum amplitude only along $y$ and none along $x$. In the intermediate state, the oscillation amplitudes are approximately equal, as expected for a non-degenerate eigenstate of the Hamiltonian in Eq.\,(\ref{Hamiltonian}) with symmetric couplings. After the second ramp, depending on whether the superexchange coupling was decreased along $x$ or $y$, we observe singlet correlations mostly along the direction of strong coupling. The measured amplitude of STO in the final state was found to be smaller than in the initial state, due to decoherence in our atomic sample which occurred on a time scale of 30~ms in our setup.  As can be seen in Fig.S2 of the supplementary material, for a total ramp time of 10\,ms (gray bar) the value of $A_\mathrm{STO}^{x}$ is $0.22(2)$, which is comparable to the STO amplitude of $0.24(2)$ obtained for the initial state $\vertvbket$ [see Fig.~\ref{Fig_VB_oscillations}(d)].

In the RVB state $\ket{\Phi_+}$, the projections on the valence bond states are given by $\mathcal{C}_x=\mathcal{C}_y=3/4$. They can be obtained from the STO amplitudes according to $\mathcal{C}_{x,y}=1/4+3/2\,A_\mathrm{STO}^{x,y}$ \cite{SOM}. By averaging the measured STO amplitudes around $J_x=J_y$, we obtain $\mathcal{C}_x=\mathcal{C}_y=0.76(7)$ [Fig.~\ref{Fig_Phi_plus}(b)], in good agreement with the theoretical prediction.
We also measured $\mathcal{C}_x$, $\mathcal{C}_y$ as a function of the coupling anisotropy $J_x/J_y$, by following the adiabatic path $\vertvbket\rightarrow\ket{\Phi_+}\rightarrow\ket{\psi(J_x/J_y)}$ with a fixed total ramp time of 10\,ms. As shown in Fig.~\ref{Fig_Phi_plus}(b), the measurement results are in good agreement with the theoretical values in the adiabatic limit (solid lines) and with a model taking into account the finite ramp time (shaded lines). 

%%%%%%%%%%%%%%%%%%%%%%%%%%%%%%%%%%%%%%%%%%%%%%%
%  Fig Phi Plus
%%%%%%%%%%%%%%%%%%%%%%%%%%%%%%%%%%%%%%%%%%%%%%%
%

\begin{figure}[t!]
\includegraphics[width=\linewidth]{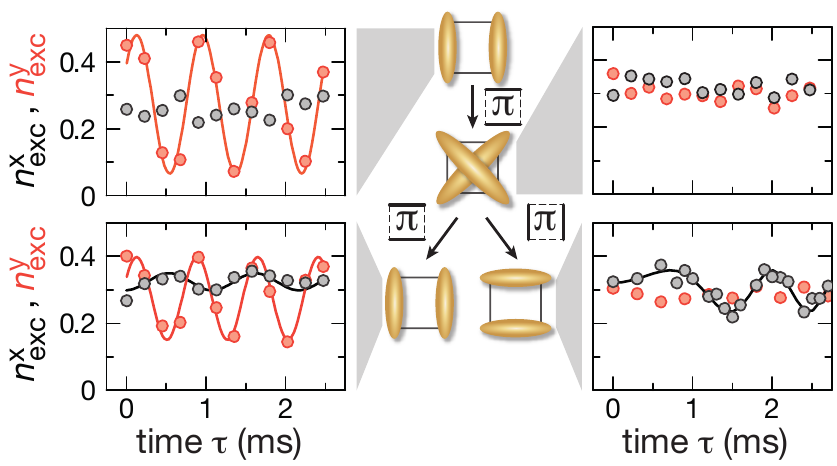}
\vspace{-0.5cm} 
\caption[]{Preparation of the $d$-wave RVB state. Schematics of the experimental sequence: starting from the state $\vertvbket$ we suddenly switch on the superexchange coupling along $x$. The $d$-wave RVB state $\left|\Phi_-\right>$ is obtained at the $\pi$ time of the subsequent periodic evolution. Measured STO are shown at the 0, $\pi$ and $2\pi$ times of the evolution. For the lower right state, we inverted the coupling direction at the $\pi$ time.
\label{Fig_Phi_minus}} 
\end{figure}

The $d$-wave RVB state $\ket{\Phi_-}$ is obtained from the state $\vertvbket$ by exchanging two spins along a bond in the $x$ direction:
\begin{equation}
	\hat{X}_{x} \vertvbket=\diagvbket.	
\end{equation}
This unitary operation was implemented by a quantum evolution of the state $\vertvbket$ under the Hamiltonian Eq.\,(\ref{Hamiltonian}) for $J_y=0$, yielding: 
\begin{equation}
\ket{\Psi(t)}=\cos\frac{\omega \tau}{2}  \, \vertvbket	
-i\sin \frac{\omega \tau}{2}  \, \diagvbket,
\end{equation}
with $\omega=2J_x/\hbar$.
For a hold time $\tau=\pi /\omega$ the initial state evolves into $\diagvbket$, characterized by $\mathcal{C}_x=\mathcal{C}_y=1/4$, and reduced STO amplitudes $A_\mathrm{STO}^{x}=A_\mathrm{STO}^{y}=1/8$. As shown in Fig.~\ref{Fig_Phi_minus}, in that state the amplitude of STO was indeed much reduced, in our case below the noise level. However, the large STO amplitude along $y$, observed both in the initial state and after one period of evolution ($\tau=2\pi /\omega$), demonstrates the coherence of the evolution and rules out a reduction of contrast at $\tau=\pi /\omega$ due to decoherence. Alternatively, after preparing the $\left|\Phi_-\right>$ state, we inverted the coupling direction by increasing in $200\;\mu$s the short-lattice depth along $x$ to $22\,E_r^s$ and decreasing the one along $y$ to $12\,E_r^s$. As shown in Fig.~\ref{Fig_Phi_minus}, we then observed a coherent evolution to a state with a large overlap with $\horvbket$, according to the measured STO.

%%%%%%%%%%%%%%%%%%%%%%%%%%%%%%%%%%%%%%%%%%%
% Fig Phi minus
%%%%%%%%%%%%%%%%%%%%%%%%%%%%%%%%%%%%%%%%%%%
%

In conclusion, we have shown direct experimental evidence of a valence-bond quantum resonance in an array of replicas of optical plaquettes, preparing and detecting minimum versions of RVB states. The $s$-wave and $d$-wave plaquette states created here could be used to encode a minimum instance of a topologically protected qubit. When stabilized by a Hamiltonian $H=J (\hat{X}_{x}+\hat{X}_{y}+\hat{X}_{xy})$, corresponding to a situation in which superexchange interaction takes also place along the diagonal bonds, these two states form a degenerate two level system which is immune to local decoherence arising, for instance, from on-site fluctuations of the external magnetic field. Such an arrangement could also be directly adapted to a setting of four coupled quantum dots to realize protected qubits in a solid state setting \cite{hanson2007}. Further extensions enabled by this work include the adiabatic connection of the plaquette RVB and valence bond solid states, or the study of their non-equilibrium dynamics upon instantaneous coupling in quantum ladders or extended two-dimensional systems. Moreover, the plaquette tools developed here could be used as building blocks for more complex protocols leading to a variety of topologically ordered states, like Laughlin states or string net condensates \cite{paredes2008minimum,paredesproc}. Finally, we note that all presented results could also be obtained using fermions instead of bosons, where the singlet valence bond is the true ground-state of a two-spin dimer. In that case, the adiabatic connection of isolated RVB states could lead to the formation of a $d$-wave superfluid upon doping \cite{altman2002plaquette,trebst2006d,rey2009controlled}.

\begin{acknowledgements}
 This work was supported by the DFG (FOR635, FOR801), the EU (STREP, NAMEQUAM, Marie Curie Fellowship to S.N.), and DARPA (OLE program). M. Aidelsburger was additionally supported by the Deutsche Telekom Stiftung.
\end{acknowledgements}

\section*{Appendix}

\bigskip

\renewcommand{\thefigure}{A\arabic{figure}}
 \setcounter{figure}{0}
\renewcommand{\theequation}{A.\arabic{equation}}
 \setcounter{equation}{0}
 \renewcommand{\thesection}{A.\Roman{section}}
\setcounter{section}{0}

\section{Filtering sequence}
\begin{figure}[h!]
\includegraphics[width=\linewidth]{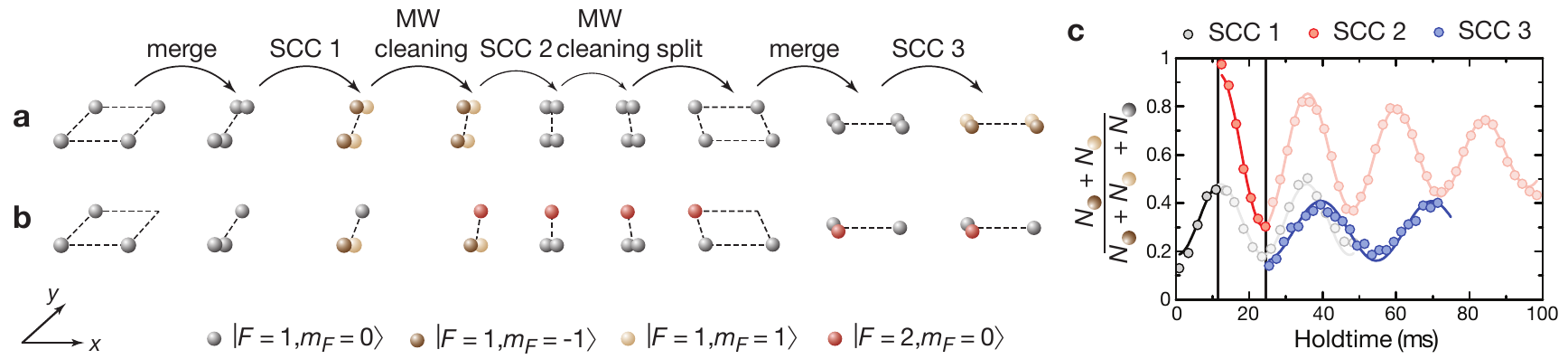}

\caption{\textbf{Schematics of the filtering sequence for plaquettes at half filling.} 
\textbf{(a), (b)} Filtering sequence applied to a plaquette initially filled with 4 atoms \textbf{(a)} and 3 atoms \textbf{(b)}. In the case of 3 atoms per plaquette no atom is eventually transferred in the Zeeman state $\left|F=1,m_F=1\right>$ that we probe at the end of the experiment. 
%The atoms in gray (blue, red, yellow) corresponds to the Zeeman state $\left|F=1,m_F=0\right>$ ($\left|F=1,m_F=1\right>$, $\left|F=1,m_F=-1\right>$, $\left|F=2,m_F=0\right>$, respectively).
\textbf{(c)} Evolution of the atom fraction in $\left|F=1,m_F=1\right>$ (within the $F=1$ manifold) during the three successive SCC used for the filtering sequence. 
\label{Fig_Filtering}} 
\end{figure}

The study presented in the main text relies on the loading of plaquettes at half filling, \emph{i.e.} with four atoms in total per plaquette. Despite the preparation of the atomic sample in a Mott insulator state at unit filling, defects are relatively likely in our system. In order to isolate the signal from correct configurations, we perform a filtering sequence that consists in transferring the atoms in plaquettes with incorrect fillings into different hyperfine states, which are not probed in the final atom imaging [see Fig.~\ref{Fig_Filtering}(a),(b)]. We first merge pairs of sites along $x$ and perform spin-changing collisions (SCC) to convert pairs of atoms in $\left|F=1,m_F=0\right>$ to the Zeeman states $\left|F=1,m_F=1\right>$ and $\left|F=1,m_F=-1\right>$. A microwave pulse then transfers the remaining atoms in $\left|F=1,m_F=0\right>$ to the state $\left|F=2,m_F=0\right>$. We then use SCC to transfer back atom pairs in $\left|F=1,m_F=0\right>$. With another microwave pulse we transfer the remaining atoms in $\left|F=1,m_F=1\right>$ due to finite SCC fidelity to $\left|F=2,m_F=2\right>$. Then we split the sites along $x$. The rest of the sequence is identical to the one described in Fig.~2(a) in the main text, \emph{i.e.} we merge pairs of sites along $y$ and perform SCC. As shown in Fig.~\ref{Fig_Filtering}, the plaquettes with one atom per site end this filtering sequence in the desired configuration. On the contrary, a hole in the initial configuration leads to the transfer of one atom to the $F=2$ hyperfine state, and finally to a configuration with one atom per site in $\left|F=1,m_F=0\right>$ before the final SCC. Therefore no atom is transferred to the state $\left|F=1,m_F=1\right>$ that we probe at the end of the experiment. Similar conclusions can be obtained for the configurations with additional holes or particles. In total about 10\% of the atoms end the filtering sequence in $\left|F=1,m_F=1\right>$ [see Fig.~\ref{Fig_Filtering}(c)].

\onecolumngrid
\vspace{1cm}
%\begin{widetext}

%\end{widetext}
%\end{minipage}
%\end{widetext}
\newpage
\twocolumngrid

\section{Measuring singlet correlations}
The quantum states prepared in this work were probed by measuring the projections $\mathcal{C}_{x,y}$ on the valence bond states along both directions. The observable $\mathcal{C}_x$ can be measured as follows: pairs of sites are merged along $x$ by decreasing the short-lattice depth along $x$ to 0 in 10\,ms. For a pair of atoms in a spin-triplet state, both atoms are transferred to the lowest Bloch band; on the contrary, due to the different parity of the quantum state, for a spin-singlet pair one expects one atom to occupy the first excited band \cite{paredes2008minimum,trotzky2010controlling}. A subsequent band-mapping technique allows one to measure the fraction $n_\mathrm{exc}^x$ of band excitations and to infer the value of $\mathcal{C}_x=2\,n_\mathrm{exc}^x$. Alternatively, the singlet correlation along $x$ can be probed through the amplitude $A_\mathrm{STO}^x$ of singlet-triplet oscillations \cite{trotzky2010controlling}. We applied a magnetic-field gradient along $x$ in order to induce a coherent oscillation between $\horsingket$ and $\hortripket$ states (here $\hortripket$ denotes a spin-triplet along the horizontal direction). An explicit calculation of the STO amplitude in the singlet state of highest energy gives $\mathcal{C}_{x,y}=1/4+3/2\,A_\mathrm{STO}^{x,y}$.

In the initial state $\vertvbket$, the measured STO amplitude  is about half of the expected value $A_\mathrm{STO}^y=1/2$. This can be attributed to residual excitations introduced by the site merging, to the residual spatial overlap after time-of-flight between atoms from the ground and first excited band, as well as to the presence of residual holes in the plaquette that do not contribute to STO.

For the measurement of $\mathcal{C}_{x,y}$ as a function of $J_x/J_y$ [see Fig.~3(b) in the main text], we followed the adiabatic path $\vertvbket\rightarrow\ket{\Phi_+}\rightarrow\ket{\psi(J_x/J_y)}$ with a fixed total ramp time of 10\,ms. The STO amplitudes were rescaled in order to give the expected value of 0.5 for the valence bond states $\horvbket$ and $\vertvbket$, using the data points at $J_x/J_y=0.006(2)$ and  $J_x/J_y=10(3)$. The rescaling factors were 3.2 and 2.3 for $\mathcal{C}_{x}$ and $\mathcal{C}_{y}$ respectively. For the point at $J_x/J_y=200(50)$ the rate of change of the couplings was the largest and adiabaticity was not maintained, as indicated by the theoretical model in Fig.~3(b) in the main text.

\section{Decoherence in the singlet subspace}
\begin{figure}[h!]
\includegraphics[width=0.95\linewidth]{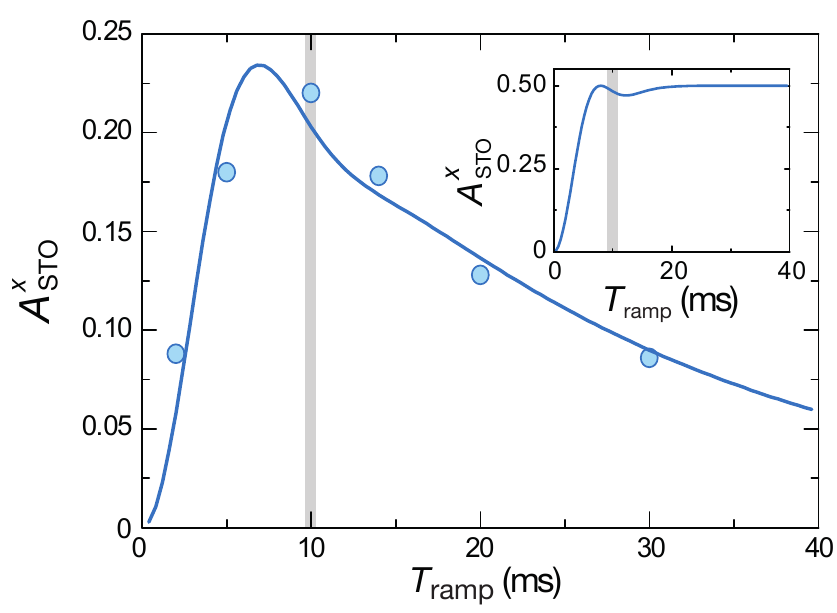}
\caption[]{\textbf{Decoherence in the singlet subspace.} 
STO amplitude after an adiabatic sweep $\vertvbket\rightarrow\left|\Phi_+\right>\rightarrow\horvbket$, as a function of the total ramp duration. The solid line is the product of the sweep fidelity expected without decoherence and of an adjustable exponential decay ($1/e$ decay time of 27(5)\,ms). The gray region indicates the ramp duration for experiment described in the main text. Inset: STO amplitude expected without decoherence, calculated by solving the Schr\"odinger equation with the Hamiltonian shown in Eq.~(1) in the main text.
\label{Fig_Decoherence}} 
\end{figure}

We investigated the decoherence of the highest energy total singlet state by measuring the STO amplitude of the adiabatic sweep $\vertvbket\rightarrow\left|\Phi_+\right>\rightarrow\horvbket$ as a function of the total ramp duration [see Fig.~\ref{Fig_Decoherence}]. The decrease in fidelity for small ramp durations is well accounted for by a numerical calculation of the evolution of the quantum state during the ramp, according to the Hamiltonian in Eq.~(1) in the main text. For the ramp duration of 10\,ms used for Fig.~3 in the main text, the calculation predicts a $A_\mathrm{STO}^x=0.48$ [see inset Fig.~\ref{Fig_Decoherence}]. The measured decrease of fidelity for longer ramp durations illustrates a decoherence mechanism in our system whose understanding would require further studies.

\vspace{2cm}

\section{Site-resolved detection}

\begin{figure}[h!]
\begin{center}
\includegraphics[width=0.9\linewidth]{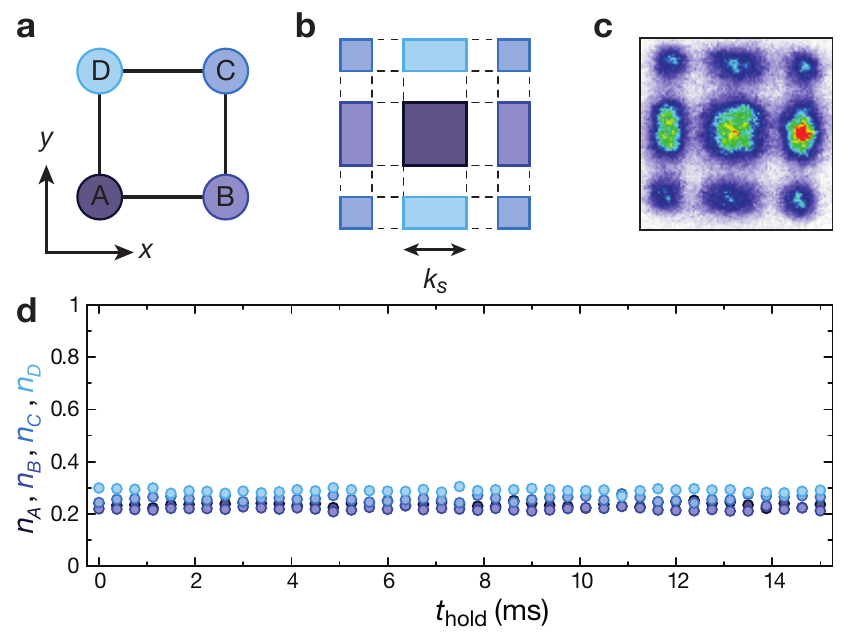}
\caption{\textbf{Site-resolved detection.}
\textbf{(a)} Schematic of the four-site plaquette.
\textbf{(b)} Brillouin zones of the 2D lattice.
\textbf{(c)} Typical momentum distribution obtained after $10\,$ms of time-of-flight.
\textbf{(d)} Measured population imbalance during the valence bond oscillations for the trace shown in Fig.~2(c) in main text.
}\label{fig:S3}
\end{center}
\end{figure}

To detect the atom numbers on the different sites of the plaquette we apply two mapping sequences along the $x$ and $y$ direction during which the populations $N_q$ are transferred to different Bloch bands analog to the technique described in Ref.~\cite{s_foelling2007direct} for isolated double wells. A subsequent band-mapping technique allows us to determine the population in the Bloch bands by counting the atom numbers in different Brillouin zones. The colors used in Fig.~\ref{fig:S3}(a) and \ref{fig:S3}(b) show the connection between the Brillouin zones and the corresponding lattice sites. A typical image obtained after $10\,$ms of time-of-flight is shown in Fig.~\ref{fig:S3}(c). The population imbalance during the valence bond oscillations is shown in Fig.~\ref{fig:S3}(d). The population in four plaquette sites remained equally populated, proving the purely spin-dynamics during the oscillation.

%\vspace{-0.3cm}

\begin{thebibliography}{25}
\bibitem [{STO()}]{STO}%
  \BibitemOpen
  \href@noop {} {}\bibinfo {note} {Under an applied magnetic field gradient, a
  triplet atom pair will evolve as
  $\ket{\psi(t)}=\ket{\uparrow\downarrow}+e^{i\delta
  t}\ket{\downarrow\uparrow}$, where $\delta$ is proportional to the gradient.
  This describes a coherent conversion of triplet pairs to singlet
  pairs.}\BibitemShut {Stop}%
\bibitem [{rec()}]{recoil}%
  \BibitemOpen
  \href@noop {} {}\bibinfo {note} {All lattice depths are given in units of the
  respective recoil energy $E_r^i=h^2/(2m\lambda_i^2)$.}\BibitemShut {Stop}%
\end{thebibliography}
\end{document}